# Do swimming animals mix the ocean?

John O. Dabiri[1,2]

1. Graduate Aerospace Laboratories, California Institute of Technology, Pasadena, CA 91125

2. Department of Mechanical and Civil Engineering, California Institute of Technology, Pasadena, CA 91125

*What began as a joke has led to new avenues of research in fluid mechanics and ocean science*

The world's oceans are in constant motion, transporting the sun's heat from the equator to the poles, bringing marine life fresh supplies of oxygen and nutrients, and sequestering nearly half of our carbon dioxide emissions since the Industrial Revolution (Sabine et al. 2004). Within this dynamic aquatic milieu exists another type of motion: the perpetual teeming of trillions of swimming animals. Are these organisms simply along for the ride, carried by the prevailing ocean currents and occasionally using their powers of locomotion to explore their surroundings; or could their propulsion result in dynamical feedbacks that influence the physical and biogeochemical structure of the ocean itself?

## *A modest proposal*

While the latter scenario might seem implausible at first, the possibility of a biogenic contribution to ocean mixing was considered by pioneering oceanographer Walter Munk in his seminal article "Abyssal Recipes" (Munk 1966). The recipe described by Munk is a steady-state balance of advection and diffusion that was shown to effectively reproduce previously measured vertical profiles of temperature, salinity, oxygen, and various chemical isotopes in the interior ocean. For example, his advective-diffusive balance of heat is given by

$$w\frac{\partial \theta}{\partial z} = \kappa \frac{\partial^2 \theta}{\partial z^2}$$

where $\theta$ is the potential temperature (i.e. temperature corrected to a standard reference pressure), $w$ is the mean flow speed in the vertical direction $z$, and $\kappa$ is the diffusivity of heat in seawater. In reality, $\kappa$ is dependent on both space and time, but Munk treats it as a constant for the purpose of estimation in the discussion that follows. The mean vertical flow $w$ arises as a consequence of a mechanism called deep water formation. As polar surface waters lose heat and form sea ice, the adjacent liquid water accumulates residual salt from the freezing process and the resulting cold, briny water sinks. This downward flux of brine and additional entrained water is balanced on average by the aforementioned upwelling at speed $w$ in the ocean interior.

Measurements of sea ice formation lead to an estimated ocean-averaged vertical upwelling speed $w$ of approximately one centimeter per day. Munk noted that the corresponding diffusivity required to balance this advection is $\kappa \approx 10^{-4} \, \mathrm{m^2 \, s^{-1}}$, a value that is three orders of magnitude greater than the molecular diffusivity of heat in water. Hence, the diffusivity arising from turbulent mixing, which can be significantly greater than molecular action, must be an essential feature in the maintenance of global ocean circulation. In his search for possible sources of energy that could account for the required turbulent diffusivity, Munk explored the potential role of animal swimming. He observed:

"The most abundant organisms in the abyssal sea, copepods, mysids, euphausiids, squid and the nektonic bristlemouths are all engaged in diurnal migration over a vertical distance of the order 1 km. This migration is associated with some transport of ocean water…"

At the time of Munk's writing, estimates for the chemical energy available to marine organisms in the ocean interior placed that total flux at 3 trillion watts. This energy flux is comparable to the total dissipation of energy by the tides, which are now an accepted driver of ocean mixing. To be sure, not all of this energy flux will contribute to animal locomotion, as other functions including metabolism and reproduction must also be supported. Nonetheless, the upper bound on the energy available for ocean mixing by swimming animals is sufficient in magnitude to make the concept initially plausible.

To estimate the turbulent diffusivity generated by those swimming animals that engage in daily cycles of diel vertical migration (DVM), Munk invoked a dimensional analysis wherein the diffusivity depends on characteristic length and time scales as $\kappa = r_1 r_2 L^2 T^{-1}$. This approximation imagines that fluid transport occurs by the animals absorbing the water properties (including temperature if it behaves as a passive scalar) at the starting location of each migration (e.g. via feeding) and subsequently delivering that fluid to the new environment at the end of the kilometer-long journey via defecation. If one takes $r_1$ to be the migrating biomass per unit of water mass, $r_2$ as the fraction of the biomass involved in diurnal feeding and defecation, and the length and time scales $L$ and $T$ to be the distance and duration of the migration, respectively, one arrives at an estimate of the diffusivity $\kappa \approx 10^{-7}$ m$^2$s$^{-1}$. This value is on par with the very slow molecular diffusivity of heat and much less than the turbulent diffusivity required for the advective-diffusive balance. In light of this result, Munk concluded that ocean mixing by swimming animals is a negligible effect.

If the concept of fluid transport via defecation strikes you as an odd proposition in the first place, it is worth noting that many years later, Munk explained to the author that this suggestion was in fact an attempt at humor (see Sidebar "Surely You're Joking!").

In the years since Munk's original recipe, an extensive body of research has established that energy inputs from wind and tidal forcing are key contributors to turbulent diffusivity in the ocean, via mixing that follows the generation and breaking of internal waves of density oscillation (Wunsch and Ferrari 2004). Subsequent work has also shown that flow over bottom topography in the ocean can generate internal waves that propagate through the interior of the ocean where wave breaking can lead to additional mixing (Waterhouse et al. 2014). Although uncertainties remain regarding the mechanisms of ocean mixing, animal swimming has generally been considered irrelevant to the process.

And yet, over the past two decades, a combination of new remote sensing tools, field measurements, and computational modeling has led to a clearer understanding of the nature and scope of biological activity in the ocean. For example, satellite-based observations of the net primary production of energy in the ocean suggest that the rate at which chemical energy is supplied to organisms in the ocean via surface photosynthesis by phytoplankton is an order magnitude greater than the value estimated by Munk, approximately 60 trillion watts (Dewar et al. 2006). Moreover, DVM has been observed in regions of the ocean that are important for global climate such as the Southern Ocean; and migrating organisms are known to reach oxygen minimum zones (Bianchi et al. 2013), where the depletion of oxygen by bacterial metabolism can limit the survivability of some organisms. This makes DVM potentially relevant as a contributor to vertical mixing that maintains the biogeochemical balance in the ocean. Finally, whereas conventional ship-based assessments of ocean biomass concluded that animals primarily occupy the sunlit surface waters of the ocean, visual and acoustic observations by remotely-piloted underwater vehicles have demonstrated that some animals—especially gelatinous zooplankton that are not easily sampled from the ocean surface—have been previously underestimated and in fact occupy the full vertical extent of the ocean (Robison 2004). The question that remains is whether the water motions induced by animal locomotion can result in turbulent diffusivities that are commensurate with significant ocean mixing.

Whereas Munk's analysis focused on fluid transported internally by swimming animals, let us consider the effect of DVM on fluid in the external vicinity of the migrating organisms. The fluid surrounding the animals will be set into motion during locomotion, with a corresponding rate of energy transfer per unit of fluid mass, $\varepsilon$, from the animals to the fluid. Famed mathematician and physicist Lewis Fry Richardson is credited with the observation that the relative dispersion associated with three-dimensional, isotropic turbulent diffusivity, $\kappa$, depends on the corresponding energy flux and characteristic length scale $L$ of the fluid motions as $\kappa \approx \varepsilon^{1/3} L^{4/3}$. Notably, theoretical models of animal swimming have predicted values of the energy flux within swimming aggregations of $\varepsilon \approx 10^{-5}$ W kg$^{-1}$ for a wide range of organisms from krill to whales (Huntley and Zhou 2004). This value is comparable to the maximum energy fluxes associated with wind and tidal forcing (Wunsch and Ferrari 2004). Field measurements of energy dissipation in the vicinity of a DVM have also recorded energy fluxes of order $10^{-5}$ W kg$^{-1}$ and even up to $10^{-4}$ W kg$^{-1}$. This result has proven difficult to reproduce, however, either because the environmental conditions of the original measurement have not been achieved in subsequent studies, or because the initial high value of recorded energy flux was an outlier (Rousseau et al. 2010).

Not all of the energy that the animals transfer to the fluid will necessarily contribute to turbulent diffusivity. The molecular viscosity of the water can dissipate some of the kinetic energy before it affects the density or chemical structure of the ocean. The efficiency of mixing will depend on the length scale $L$ of the fluid motions, not only through Richardson's 4/3-power law, but also in comparison to the characteristic length scale of the density stratification in the ocean. The buoyancy length scale $L_B$ that quantifies the stratification of fluid density $\rho$ is given by dimensional analysis as $L_B \approx \varepsilon^{1/2} N^{-3/2}$, where $N = \sqrt{\dfrac{-g}{\rho} \dfrac{\partial \rho}{\partial z}}$ is called the buoyancy frequency and is a measure of the strength of the stratification. The action of turbulent diffusivity $\kappa$ on a fluid with buoyancy frequency $N$ leads to a buoyancy flux (i.e., a rate of increase in potential energy) of $\kappa N^2$. Hence, the efficiency of the mixing process can be quantified by the ratio $\kappa N^2 / \varepsilon$. Based on the foregoing scaling relationships, this mixing efficiency is related to the characteristic length scales as

$$\frac{\kappa N^2}{\varepsilon} = \left(\frac{L}{L_B}\right)^{4/3}.$$

This relationship holds until turbulent motion (and the corresponding length scale $L$) is suppressed by the stratification, resulting in a maximum mixing efficiency less than 100 percent.

Given that typical buoyancy length scales in the ocean based on background dissipation and stratification are of order $L_B \approx 10^{-1}$ to 10 m, the question of whether swimming animals can mix the ocean ultimately becomes a question of whether the organisms, often only millimeters to centimeters in size, can generate coherent fluid motions with a vertical extent (i.e. parallel to the density gradient) that is at least an order of magnitude larger than their individual bodies.

The past several years have seen an increasing number of studies aiming to address this question. The most straightforward approach is to assume that the primary fluid motions associated with animal swimming are the swirling vortices created by the flapping, paddling, or jetting motions of the organism appendages. In this case, the characteristic length scale of fluid motion $L$ is expected to be of the same order of magnitude as the body size, which is typically much smaller than the buoyancy length scale $L_B$. It is on this basis that some have argued that the energy input to the ocean by swimming animals, however immense, is almost entirely dissipated as heat by the action of viscosity before it can contribute to turbulent diffusivity (Visser 2007; Kunze 2019). Field measurements of fish schools in the ocean have supported this notion (Pujiani et al. 2015).

*Global swarming*

If we return to Munk's picture of DVM, we find that there exists another length scale in addition to the individual animal size and the vertical migration distance that is potentially relevant to the biogenic mixing process. As the animals migrate vertically, they form swarms that can be tens of meters in vertical extent (Sato et al. 2013). Could the dynamics at this scale play a role in biogenic mixing?

To answer this question, researchers at Caltech devised a laboratory experiment in which aggregations of brine shrimp (the "sea monkeys" from your local pet store) were coaxed into simulating DVM under controlled conditions (Wilhelmus and Dabiri 2014). Because the organisms are attracted to blue light, the vertical translation of a $4.47 \times 10^{-7}$ m wavelength laser beam caused the aggregation to give chase, resulting in on-demand DVM (Figure 1). Concurrent measurements of the flow around the animals revealed that the induced fluid motions from each of the organisms coalesced to generate a coherent jet directed opposite to the direction of migration. That jet broke down into swirling eddies that were individually several times the size of any individual animal, suggesting that mixing could potentially occur also occur at length scales larger than the individual animals.

The aforementioned laboratory experiments were limited in some important ways. First, the field of view of the flow measurements was constrained to a 3-cm x 3-cm square in order to quantitatively resolve the flow features in the aggregation. Hence, the full extent of the induced jet flow could not be determined. Second, the experiments were conducted in water with constant density, unlike the density-stratified structure of the ocean. The effect of turbulence-suppressing stratification was therefore absent. Finally, although the measurements indicated the transient formation of large-scale eddies, irreversible mixing of the water column, i.e., a permanent change in the structure of the water column after those eddy motions ceased, was not demonstrated.

These limitations were addressed in a subsequent series of laboratory experiments (Houghton et al. 2018; Houghton and Dabiri 2019). In addition to incorporating both laser- and LED-based visual stimuli, the facility also enabled experiments in a density-stratified and/or oxygen-stratified water column; flow visualizations with a field of view up to a half-meter across; and measurements of irreversible mixing. The results showed that for vertical migration over time scales similar to those in the ocean, the animals mix the water column at rates up to three order of magnitude larger than molecular diffusion (Figure 2). Moreover, multi-scale flow visualizations revealed that the induced jet observed in earlier experiments is propelled through the entire vertical extent of the aggregation (Figure 3). In the laboratory, that corresponded to a length scale of almost 50 centimeters, as compared to the 1-cm length of the individual animals. Indeed, this dramatic jet formation and its subsequent breakdown are sufficient to account for the three orders of magnitude enhancement in diffusivity that was observed in the irreversible mixing experiments.

Whereas the laboratory experiments were conducted using gradients in salt concentration to achieve density stratification, much of the ocean is density-stratified due to thermal gradients. The irreversible mixing of thermal stratification could be even more significant than the measured salt mixing, because the more rapid molecular diffusion of heat relative to salt will prevent restratification of fluid displaced by the eddying motions (Nash and Moum 2002; Shih et al. 2005; Jackson and Rehmann 2014). In one set of experiments comparing mixing of thermal and salinity stratification, the thermal mixing was up to 30 times that of salt mixing for the same forcing (Jackson and Rehmann 2014). Translated to the present biogenic mixing results, that would suggest turbulent diffusivities up to $10^{-5}$ $m^2s^{-1}$, which is comparable to median values from ocean microstructure measurements (Monismith et al. 2018).

***Fishing for clues***

The foregoing discussion informs consideration of whether swimming animals *could* mix the ocean, i.e., whether it is physically possible or not. However, proof that they actually *do* make a significant contribution to turbulent diffusivity in the ocean does not yet exist, and that proof will require in situ observations of physical impacts that can be unambiguously attributed to animal swimming. The best efforts to date have relied on fortuitous encounters with aggregations of swimming animals, work that has incidentally been recognized with the 2023 Ig Nobel Prize in Physics (Fernandez Castro et al. 2022). However, those studies are difficult to generalize, since in some cases the animals were not migrating vertically and in each case the primary focus has been on the energy dissipation that could be correlated with the presence of animals.

Further field work has reported the first direct observation of vertical velocity 'anomalies' associated with aggregations of krill in the ocean (Tarling and Thorpe 2017). The magnitude of the measured vertical velocities is consistent with those observed in the laboratory, providing a hint that the effects measured in the laboratory might be representative of analogous phenomena in the ocean. If the downward jet flow created by the animals in the lab is similarly limited only by the vertical extent of DVM in the ocean, those fluid dynamic features would exhibit length scales of tens of meters, which would be sufficient to affect ocean mixing.

It should be noted that an implicit assumption in previous searches for a hydrodynamic signature of ocean mixing by swimming animals has been that the mixing must be spatially coincident with the aggregation. Yet, the available laboratory observations suggest that the flow instabilities that lead to mixing may occur adjacent to the aggregation rather than within it. The Kelvin-Helmholtz instability observed at the edges of the coalesced jets created by the migrating brine shrimp aggregation is one such example. In the case of a Rayleigh-Taylor instability generated by DVM, restratification of the vertically displaced fluid mass could trigger internal waves of density oscillation that radiate away from the aggregation and do not mix via wave breaking until far removed from the animals. These physics, if present, could complicate efforts to empirically connect animal swimming to associated mixing events. Unlike wind forcing, which can be observed globally via remote sensing, or tidal forcing, which has a predictable cycle, the precise location of DVM events exhibits stochasticity that limits their accessibility.

To address this challenge of identifying DVM in the field, new concepts are emerging to indirectly detect biogenic ocean mixing based on associated physics or biology. These include strategies that would leverage the magnetic signature of vertical flow currents in Earth's geomagnetic field (Fu and Dabiri 2023), or exploit sampling of environmental DNA associated with DVM (Office of Naval Research 2022). Both of these approaches could circumvent challenges associated with uncertainties in identifying swimming aggregations in field studies, while also avoiding triggering escape or avoidance behaviors that can cause organisms to avoid measurement instrumentation deployed in their vicinity.

In light of the myriad challenges associated with field measurements of biogenic ocean mixing, the most immediate way forward in the search for answers to the headline question will likely involve further laboratory and computational studies to enable anticipation of the fluid transport processes most likely to be associated with the swimming of different animal species in the ocean. The physics of internal wave generation by DVM is an open question and possibly ripe for theoretical solution. Field research in smaller bodies of water such as lakes can also constrain the problem, both conceptually and literally, as the migrating animals are often located more easily in those cases. The 'smoking gun' of ocean mixing by swimming animals may ultimately lie not in hydrodynamic signatures, but in the biogeochemical consequences of those fluid mechanical effects. Restructured profiles of oxygen, carbon, nutrients, or bacterial species composition in the ocean due to animal swimming could persist long after the fluid motions have faded away. And these effects can have important consequences for the ocean, regardless of whether they occur globally or only on local scales.

## Acknowledgements

Many colleagues have contributed to our current understanding of the headline question, and not all have been referenced here due to space constraints. I want to especially thank Isabel Houghton, Monica Martinez Wilhelmus, Kakani Katija, Eckart Meiburg, Jeffrey Koseff, and Stephen Monismith for key insights along the way, and an engaged community of oceanographers for constructive and often spirited feedback.

## References

Bianchi, D., E. D. Gilbraith, D. A. Carozza, K. A. S. Mislan, and C. A. Stock. 2013. Nature Geosci. 6: 545-548.

Dewar, W. K., R. J. Bingham, R. L. Iverson, D. P. Nowacek, L. C. St. Laurent, and P. H. Wiebe. 2006. J. Marine Res. 64: 541-561.

Fernandez Castro, B., M. Pena, E. Nogueira, and others. 2022. Nature Geosci. 15: 287-292.

Fu, M. K., and J. O. Dabiri. 2023. Magnetic signature of vertically migrating aggregations in the ocean. Geophys. Res. Lett. 50: e2022GL101441.

Houghton, I. A., J. R. Koseff, S. G. Monismith, and J. O. Dabiri. 2018. Vertically migrating swimmers generate aggregation-scale eddies in a stratified column. Nature 556: 497-500.

Houghton, I. A., and J. O. Dabiri. 2019. Alleviation of hypoxia by biologically generated mixing in a stratified water column. Limnol. Oceanogr. 64: 2161-2171.

Huntley, M. E., and M. Zhou. 2004. Influence of animals on turbulence in the sea. Marine Ecol. Prog. Ser. 273: 65-79.

Jackson, P. R., and C. R. Rehmann. 2014. Experiments on differential scalar mixing in turbulence in a sheared, stratified flow. J. Phys. Oceanogr. 44: 2661-2680.

Kunze, E. 2019. Biologically generated mixing in the ocean. Annu. Rev. Mar. Sci. 11: 215-226.

Monismith, S. G., J. R. Koseff, and B. L. White. 2018. Mixing efficiency in the presence of stratification: when is it constant? Geophys. Res. Lett. 5: 5627-5634.

Munk, W. 1966. Abyssal recipes. Deep-Sea Res. 13: 707-730.

Nash, J. D., and J. N. Moum. 2002. Microstructure estimates of turbulent salinity flux and the dissipation spectrum of salinity. J. Phys. Oceanogr. 32: 2312-2333.

Office of Naval Research Announcement N00014-22-S-F002, Topic 15. 2022.

Pujiani, K., J. N. Moum, W. D. Smyth, and S. J. Warner. 2015. Distinguishing ichthyogenic turbulence from geophysical turbulence. J. Geophys. Res. 120: 3792-3804.

Robison, B. H. 2004. Deep pelagic biology. J. Exp. Marine Biol. Ecol. 300: 253-272.

Rousseau, S., E. Kunze, R. Dewey, K. Bartlett, and J. Dower. 2010. On turbulence production by swimming marine organisms in the open ocean and coastal waters. J. Phys. Ocean. 40: 2107-2121.

Sabine, C. L., R. A. Feely, N. Gruber, and others. 2004. The ocean sink for anthropogenic $CO_2$. Science 305: 367-371.

Sato, M., J. F. Dower, E. Kunze, and R. Dewey. 2013. Second-order seasonal variability in diel vertical migration timing of euphausiids in a coastal inlet. Mar. Ecol. Prog. Ser. 480: 39-56.

Shih, L. H., J. R. Koseff, G. N. Ivey, and J. H. Ferziger. 2005. Parameterization of turbulent fluxes and scales using homogeneous sheared stably stratified turbulence simulations. J. Fluid Mech. 525: 193-214.

Tarling, G. A., and S. E. Thorpe. 2017. Oceanic swarms of Antarctic krill perform satiation sinking. Proc. Roy. Soc. B 284: 20172015.

Visser, A. W. 2007. Ocean science. Biomixing of the oceans? Science 316: 838-839.

Waterhouse, A. F., J. A. MacKinnon, J. D. Nash, and others. 2014. Global patterns of diapycnal mixing from measurements of the turbulent dissipation rate. J. Phys. Oceanogr. 44: 1854-1872.

Wilhelmus, M. M., and J. O. Dabiri. 2014. Observations of large-scale fluid transport by laser-guided plankton aggregations. Phys. Fluids 26: 101302.

Wunsch, R., and R. Ferrari. 2004. Vertical mixing, energy, and the general circulation of the oceans. Annu. Rev. Fluid Mech. 36: 281-314.

**Sidebar: Surely You're Joking!**

At the beginning of the author's interest in this research topic, Walter Munk provided his perspective on ocean mixing by swimming animals, as well as sage advice for a young assistant professor. That correspondence is excerpted below:

```
Date: Tue, 27 Feb 2007 18:07:35 -0800 (PST)
From: John O. Dabiri <jodabiri@its.caltech.edu>
To: wmunk@ucsd.edu
Subject: ocean mixing by animals?
```

```
Dear Dr. Munk,
```

```
I have recently read many of your papers with keen interest, especially
your 1998 paper in Deep-Sea Research that provides an excellent outline of
the state of ocean research for those of us who are not native to the field.
```

```
I am writing to get your opinion on a growing school of thought that
ascribes to aquatic animals a non-negligible role in mixing the ocean. I
understand that this is not a popular idea, yet I have read a few recent
```

papers that make a reasonable case. In particular, they note that
order-of-magnitude estimates based on propulsive power do not place the
energy input from animals out of the realm of possible contributions to ocean
mixing. Also, while the animals are not found at most 'abyssal' depths, one
could similarly argue that the energy input from the wind occurs at
shallow depths and still effect abyssal mixing. Yet, before pursuing the
topic of biological ocean mixing any further, I thought I would seek the
perspective of someone who has been thinking about this problem longer and
more deeply than I have.

I would be grateful for any comments you would be willing to offer.

With best regards,

John
<><

**Date: Wed, 28 Feb 2007 18:37:41 -0800**
**From: Walter Munk <wmunk@ucsd.edu>**
**To: John O. Dabiri <jodabiri@its.caltech.edu>**
**Subject: Re: ocean mixing by animals?**

Dear John:

It was partly an attempt at humor when I suggested many years ago
that diurnal migration could lead to appreciable mixing.  And I was
amazed at recent papers who think this is not a joke.

And people thought it was a lunatic idea when Carl Wunsch and I
suggested that the moon (via lunar tides) could have anything to do
with mixing. And now that is generally accepted.  I must say that we
preferred the era when we were called lunatic to the era where people say "we all know".

I can only say:  work on problems you most enjoy. Strange things can happen under way.

Walter

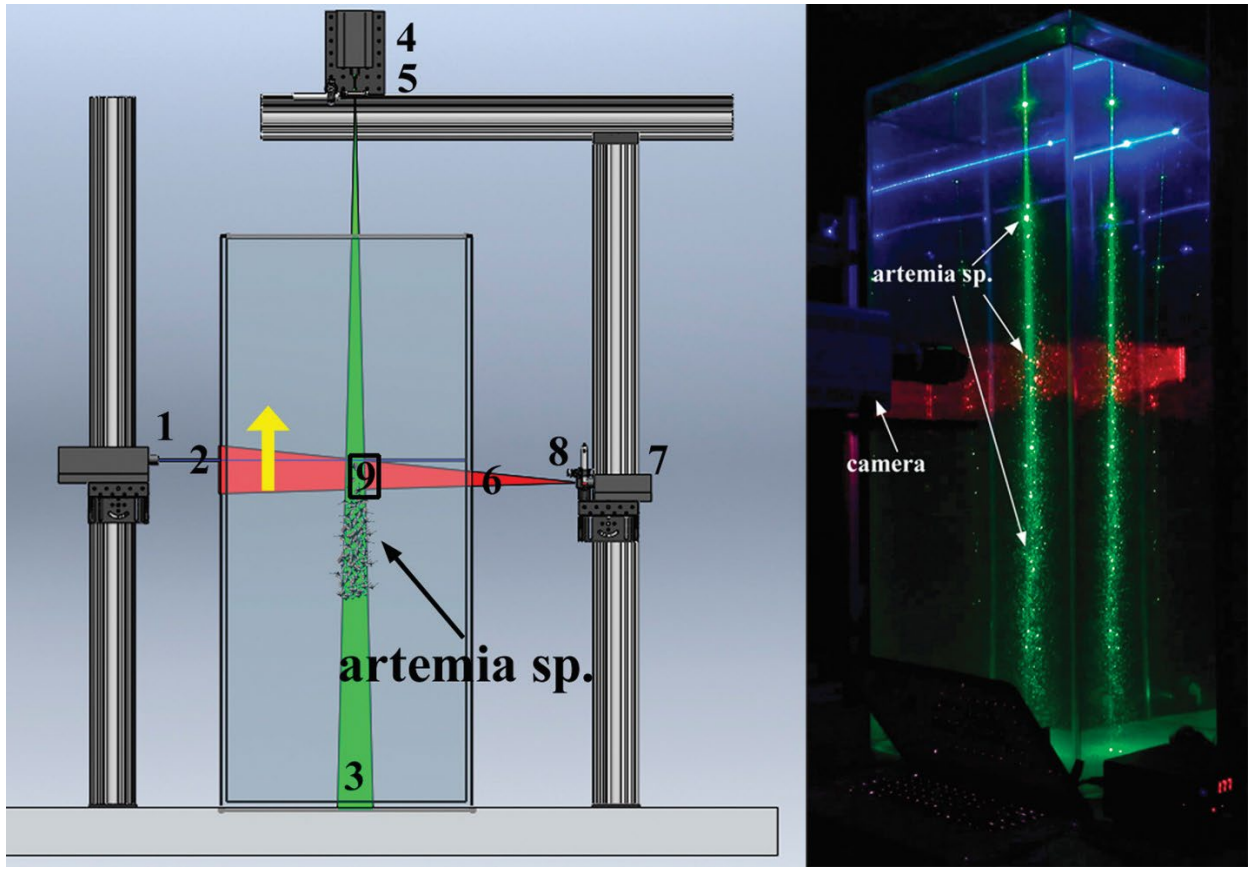

**Figure 1.** A laboratory facility creates on-demand DVM of brine shrimp (*Artemia sp.*) using laser guidance. The vertically translating laser (1) emits a continuous beam of blue light (2) that passes horizontally through the water. Animals chase the blue light within a narrow, green light sheet (3) created by a second laser (4) and optics (5). A third, vertically oriented laser sheet (6) created by a red laser (7) and optics (8) enables tracking of suspended particles in the water for flow measurements. Right panel: Actual experiment as the blue laser arrives at the top of the tank (double image is due to corner reflection). Adapted from Wilhelmus and Dabiri (2014).

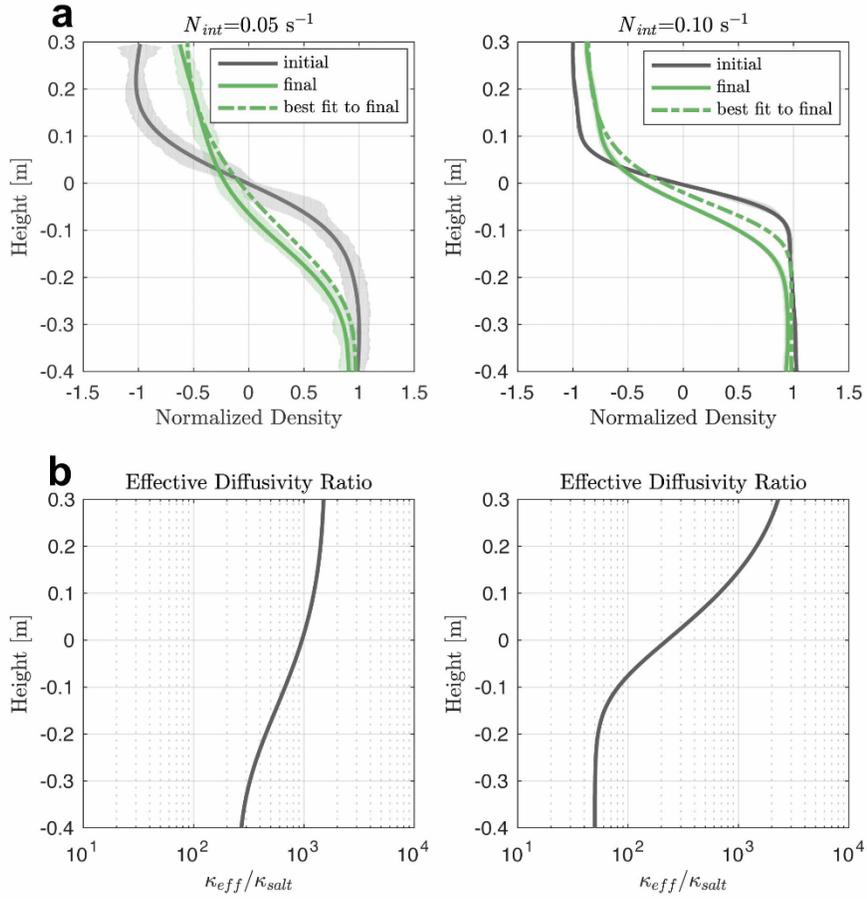

**Figure 2.** Effective diffusivity due to vertical migration. Top: Initial (black) and final (solid green) density profiles for background stratifications of $N = 0.05$ s$^{-1}$ (left) and $N = 0.10$ s$^{-1}$ (right). A numerically calculated, best-fit error function density profile is shown in dashed green. Bottom: Ratio of effective diffusivity to salt molecular diffusivity as a function of height for each experiment above, based on the best-fit density profiles. Adapted from Houghton et al. (2018).

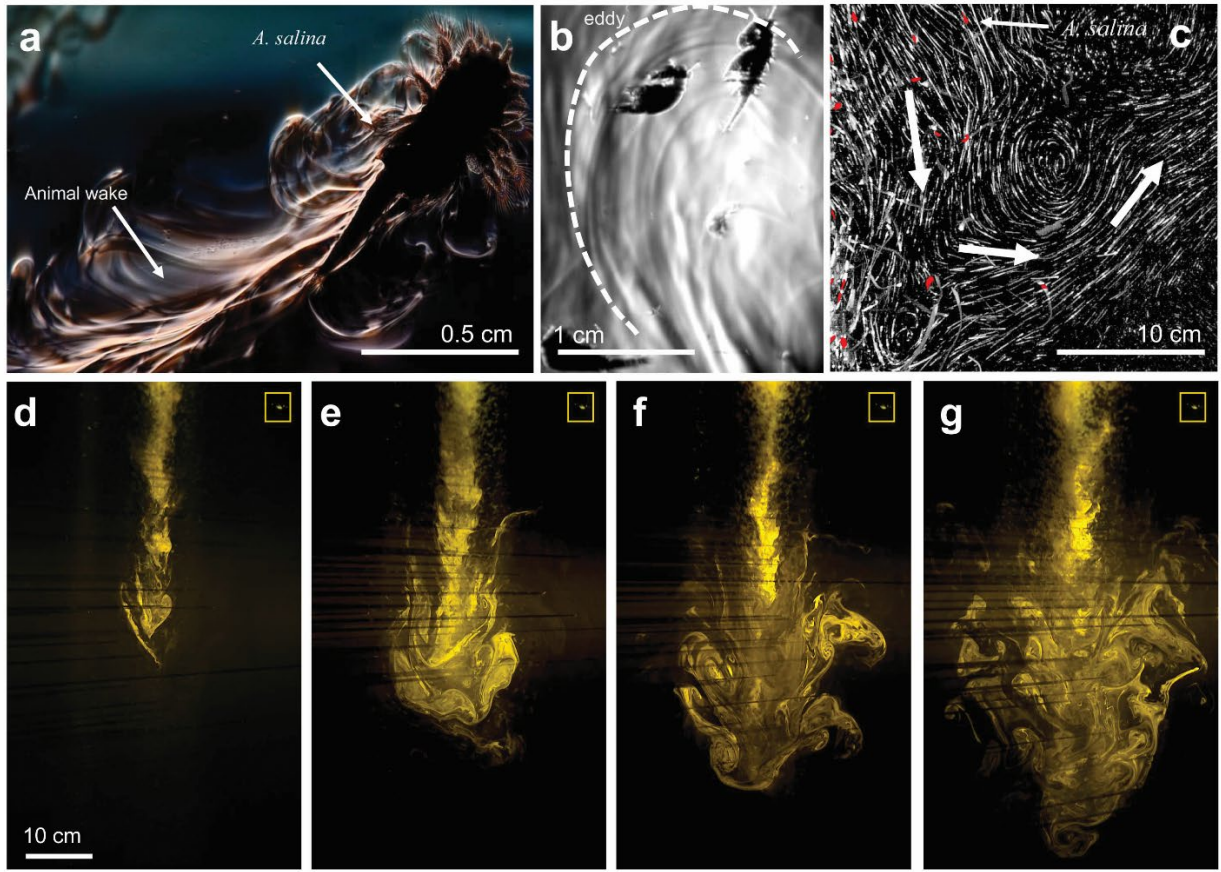

**Figure 3.** Flow visualization from animal to aggregation scales. (a) Single animal schlieren, conducted with the assistance and facilities of Dr. Rudi Strickler (U. Wisconsin-Milwaukee), showing fluid motion in the wake of a single animal. (b) Intermediate-scale schlieren imaging with a 5 cm field of view centered 8 cm to the left of the migration, showing the laterally propagating eddies perturbing the stable background density. (c) Pathlines of 10 micrometer neutrally buoyant particles illustrate the fluid motion to the right of the migration, including the downward jet proximate to the aggregation and eddy motion on the periphery. Individual swimmers are overlaid in red. White arrows show flow direction. (d)-(g) Planar cross-section of laser-induced fluorescence of a tracer dye propelled downward through the extent of the migration at 54 s, 122 s, 185 s, and 292 s after the beginning of a vertical migration, respectively.

Surface fluid is propelled downward through a stable stratification over 50 cm vertically, with large-scale flow structures entraining fluid proximate to the downward jet. Animals in the laser sheet cast horizontal shadows through the illuminated dye. Individual animal size is highlighted in the upper right hand corner inset. Adapted from Houghton et al. (2018).